\documentclass[12pt]{article}
\usepackage{amssymb}

\usepackage{amsmath}

\newtheorem{theorem}{Theorem}
\newtheorem{acknowledgement}[theorem]{Acknowledgement}

\newtheorem{remark}[theorem]{Remark}

\newenvironment{proof}[1][Proof]{\textbf{#1.} }{\ \rule{0.5em}{0.5em}}
\input{tcilatex}

\begin{document}

\title{Integrable hydrodynamic chains}
\author{Maxim V.Pavlov}
\maketitle

\begin{abstract}
A new approach for derivation of Benney-like momentum chains and integrable
hydrodynamic type systems is presented. New integrable hydrodynamic chains
are constructed, all their reductions are described and integrated. New
(2+1) integrable hydrodynamic type systems are found.
\end{abstract}


keywords: Euler-Darboux-Poisson equation, hydrodynamic type system, momentum
chain, conservation law, commuting flow, integrable reduction.

\section{Introduction}

The integrable hydrodynamic chain%
\begin{equation}
\partial _{t}A_{k}=\partial _{x}A_{k+1}+kA_{k-1}A_{0,x}\text{, \ \ \ \ \ \ \
}k=0\text{, }1\text{, ...}  \label{1}
\end{equation}%
for the first time was introduced by D.J. Benney in a theory of finite-depth
fluid (see \cite{Benney}). Here moments $A_{k}$ are infinite number of field
variables. Later, it was shown that these moments satisfy a dispersionless
limit of KP hierarchy determined by the Sato pseudo-differential operator%
\begin{equation*}
\hat{L}=\partial _{x}+A_{0}\partial _{x}^{-1}+A_{1}\partial _{x}^{-2}+...,
\end{equation*}%
which in dispersionless limit is reduced to%
\begin{equation}
\lambda =\mu +\frac{A_{0}}{\mu }+\frac{A_{1}}{\mu ^{2}}+...  \label{2}
\end{equation}%
The Benney momentum chain can be written in equivalent form (see \cite{Gib})%
\begin{equation}
\lambda _{t}-\mu \lambda _{x}=\frac{\partial \lambda }{\partial \mu }[\mu
_{t}-\partial _{x}(\frac{\mu ^{2}}{2}+A_{0})].  \label{3}
\end{equation}%
If $\lambda =\func{const}$, then $\mu $ is a generating function (with
respect to the parameter $\lambda $) of the conservation law densities%
\begin{equation}
\mu _{t}=\partial _{x}(\frac{\mu ^{2}}{2}+A_{0}).  \label{4}
\end{equation}%
In the case, when several first moments $A_{k}$ are functionally-independent
($k=0$, $1$, ..., $N-1$), the corresponding hydrodynamic reductions
(''hydrodynamic'' reduction means that all higher moments $A_{k}$ ($k=1$, $2$%
, ...) could not depend of any derivatives of lower moments $A_{k}$ ($k=0$, $%
1$, ..., $N-1$); in opposite case, such reductions one can call as
''differential'' reductions. In this article we concentrate attention at
hydrodynamic reductions only in spirit of \cite{Gibbons}) are the
hydrodynamic type systems written in Riemann invariants $r^{i}$%
\begin{equation}
r_{t}^{i}=\mu _{i}(\mathbf{r})r_{x}^{i}\text{, \ \ \ \ \ \ }i=1\text{, }2%
\text{, ..., }N,  \label{5}
\end{equation}%
i.e. hydrodynamic type system has diagonal form in these field variables and
there is no summation over each repeated index, see for instance \cite{Tsar}%
. The Riemann invariants $r^{i}$ and the characteristic velocities $\mu _{i}(%
\mathbf{r})$ are determined by conditions%
\begin{equation*}
r^{i}=\mu _{i}+\frac{A_{0}}{\mu _{i}}+\frac{A_{1}}{\mu _{i}^{2}}+\frac{A_{2}%
}{\mu _{i}^{3}}+...\text{, \ \ \ \ \ \ \ \ \ \ \ }1=\frac{A_{0}}{\mu _{i}^{2}%
}+2\frac{A_{1}}{\mu _{i}^{3}}+3\frac{A_{2}}{\mu _{i}^{4}}+...
\end{equation*}%
(see (\ref{2}) and (\ref{3})). These hydrodynamic type systems are
integrable too (all moments $A_{k}$ are some functions of $r^{i}$, which are
determined by compatibility conditions with whole Benney momentum chain, see %
\cite{Gibbons}). These hydrodynamic type systems have the same generating
functions of conservation laws (see (\ref{4})) and the commuting flows (see %
\cite{Maks2}; ''commuting flows'' means, that the Riemann invariants $r^{i}$
simultaneously are functions of infinite number of independent variables $%
t_{k}$, $k=0$, $1$, ... , here $t_{0}\equiv x$, $t_{1}\equiv t$.)%
\begin{equation}
\mu (\lambda )_{\tau (\tilde{\lambda})}=\partial _{x}\ln [\mu (\lambda )-\mu
(\tilde{\lambda})],  \label{6}
\end{equation}%
where%
\begin{equation*}
\partial _{\tau (\tilde{\lambda})}=\partial _{t_{0}}+\frac{1}{\tilde{\lambda}%
}\partial _{t_{1}}+\frac{1}{\tilde{\lambda}^{2}}\partial _{t_{2}}+...
\end{equation*}%
Moreover, the generating function (with respect to the parameter $\tilde{%
\lambda}$) of solutions for any reduction (\ref{5}) can be found by the
Tsarev generalized hodograph method (\cite{Tsar}, also see \cite{Maks2})%
\begin{equation}
x+\mu _{i}(\mathbf{r})t=\frac{1}{\mu _{i}(\mathbf{r})-\mu (\tilde{\lambda})}%
\text{, \ \ \ \ \ \ }i=1\text{, }2\text{, ..., }N\text{ }  \label{7}
\end{equation}%
Thus, if some hydrodynamic type system is recognized as a reduction of the
Benney momentum chain, it means that this system has most properties of the
Benney momentum chain.

The idea presented in this paper is the following: if one can introduce the
moments $A_{k}$ for given integrable hydrodynamic type system (\ref{5}),
then \textbf{one can ignore the origin} (i.e. given hydrodynamic type
system) of this \textit{hydrodynamic chain}%
\begin{equation*}
\partial _{t}A=F(\mathbf{A})A_{x},
\end{equation*}%
where $A$ is an infinite-number component vector, $F(\mathbf{A})$ is an
infinite-number component matrix. The next step is a description of all
possible integrable hydrodynamic reductions (one of them, of course, must be
the original hydrodynamic type system (\ref{5}))%
\begin{equation}
r_{t}^{i}=V_{i}(\mathbf{r})r_{x}^{i}\text{, \ \ \ \ \ \ \ }i=1\text{, }2%
\text{, ..., }M,  \label{red}
\end{equation}%
where $M$ is not connected with $N$ (see (\ref{5})), and $V_{i}$ satisfy
some nonlinear system of PDE's (see below). Also, we assume that $V_{i}\neq
V_{k}$ for any $i\neq k$ (this is a necessary condition for the application
of the Tsarev generalized hodograph method). Thus, every hydrodynamic chain
constructed in this way can be regarded as a \textit{huge} \textit{box} for
some variety of the integrable hydrodynamic type systems.

For instance, the particular case of gas dynamics%
\begin{equation}
u_{t}=\partial _{x}[\frac{u^{2}}{2}+\frac{\eta ^{\gamma -1}}{\gamma -1}]%
\text{, \ \ \ \ }\eta _{t}=\partial _{x}(u\eta ),  \label{gas}
\end{equation}%
for $\gamma =2$ (shallow water equations):%
\begin{equation}
u_{t}=\partial _{x}[\frac{u^{2}}{2}+\eta ]\text{, \ \ \ \ }\eta
_{t}=\partial _{x}(u\eta )  \label{shallow}
\end{equation}%
satisfies for Benney momentum chain (\ref{1}) if one introduces the moments $%
A_{k}=u^{k}\eta $.

It is easy to check, that the Benney momentum chain has a more general (the
Zakharov) reduction $A_{k}=\overset{N}{\underset{i=1}{\sum }}u_{i}^{k}\eta
_{i}$ (see \cite{Zakh}), which create a dispersionless limit of vector
nonlinear Shrodinger equation (VNLS)%
\begin{equation}
\partial _{t}u_{i}=\partial _{x}[\frac{u_{i}^{2}}{2}+\overset{N}{\underset{%
k=1}{\sum }}\eta _{k}]\text{, \ \ \ \ \ \ }\partial _{t}\eta _{i}=\partial
_{x}(u_{i}\eta _{i})\text{, \ \ \ \ \ \ }i=1\text{, }2\text{, ..., }N
\label{VNLS}
\end{equation}%
\qquad \qquad

\begin{remark}
Infinite series (\ref{2}) under this Zakharov reduction yields a more
compact expression (see \cite{Zakh})%
\begin{equation}
\lambda =\mu +\overset{N}{\underset{k=1}{\sum }}\frac{\eta _{k}}{\mu -u_{k}}.
\label{Rim}
\end{equation}%
It is easy to check that the dispersionless limit of VNLS satisfies equation
(\ref{3}) with respect to equation of Riemann surface (\ref{Rim}).
\end{remark}

Obviously, in both above-mentioned cases, corresponding hydrodynamic type
systems (\ref{shallow}) and (\ref{VNLS}) have the same generating functions
of conservation law densities (\ref{4}) and commuting flows (\ref{6}) as
whole Benney momentum chain (\ref{1}). Here we demonstrate this approach on
an example of a new hydrodynamic chain, which contains some important
reductions well known in mathematics, fluid dynamics, nonlinear optics,
biology and chemistry.

The main classification problem in the theory of integrable hydrodynamic
type systems can be re-formulated as the problem of description of all
possible integrable hydrodynamic chains. For the simplicity, any hierarchy
of the hydrodynamic chains can be written in a conservative form (see, for
instance, (\ref{four}) below)%
\begin{equation*}
\partial _{t_{n}}A_{k}=\partial _{x}F_{k}(A_{k+n}\text{, }A_{k+n-1}\text{,
... , }A_{0})\text{, \ \ \ \ \ \ }k\text{, }n=0\text{, }1\text{, ...}
\end{equation*}%
If one can classify all possible functions $F_{k}$, it would mean that all
the hydrodynamic type systems embedded in such hydrodynamic chains by
different reductions are classified too. In simplest case ($N=2$)%
\begin{equation*}
\partial _{t}A_{0}=\partial _{x}F_{0}(A_{0}\text{, }A_{1})\text{, \ \ \ \ \
\ }\partial _{t}A_{1}=\partial _{x}F_{1}(A_{0}\text{, }A_{1}\text{, }A_{2})%
\text{,}
\end{equation*}%
where $A_{2}(A_{0}$, $A_{1})$ is conservation law density of reduced
hydrodynamic type system. Then governing equation for function $A_{2}$ is
quasilinear%
\begin{equation*}
f_{\upsilon }w_{uu}=[f_{u}-\varphi _{\upsilon }-\varphi _{w}w_{\upsilon
}]w_{u\upsilon }+[\varphi _{u}+\varphi _{w}w_{u}]w_{\upsilon \upsilon },
\end{equation*}%
where%
\begin{equation*}
u\equiv A_{0}\text{, \ \ \ }\upsilon \equiv A_{1}\text{, \ \ \ }w\equiv A_{2}%
\text{, \ \ \ \ }f\equiv F_{0}\text{, \ \ \ \ }\varphi \equiv F_{1}.
\end{equation*}%
When $F_{0}=\upsilon $, \ $F_{1}=w(u$, $\upsilon )-u^{2}/2$, this is $2-$%
component reduction of the Benney momentum chain (see \cite{Gibbons})%
\begin{equation*}
w_{uu}=-w_{\upsilon }w_{u\upsilon }+(w_{u}-u)w_{\upsilon \upsilon };
\end{equation*}%
if $F_{0}=\upsilon -u^{2}$, $F_{1}=w(u$, $\upsilon )-u\upsilon $, then
corresponding equation%
\begin{equation*}
w_{uu}=-(u+w_{\upsilon })w_{u\upsilon }+(w_{u}-\upsilon )w_{\upsilon
\upsilon },
\end{equation*}%
can be solved in parametric form%
\begin{equation*}
\begin{array}{c}
w=\frac{1}{6}[A^{\prime \prime }(s)+B^{\prime \prime }(r)]^{3}+[A^{\prime
\prime }(s)+B^{\prime \prime }(r)][A^{\prime }(s)-sA^{\prime \prime
}(s)+B^{\prime }(r)-rB^{\prime \prime }(r)]+ \\
s^{2}A^{\prime \prime }(s)-2sA^{\prime }(s)+2A(s)+r^{2}B^{\prime \prime
}(r)-2rB^{\prime }(r)+2B(r), \\
\\
\upsilon =\frac{1}{2}[A^{\prime \prime }(s)+B^{\prime \prime
}(r)]^{2}+A^{\prime }(s)-sA^{\prime \prime }(s)+B^{\prime }(r)-rB^{\prime
\prime }(r)\text{, \ \ \ \ }u=A^{\prime \prime }(s)+B^{\prime \prime }(r),%
\end{array}%
\end{equation*}%
where $A(s)$ and $B(r)$ are arbitrary functions. Thus, $2-$component reduced
(hydrodynamic type) system in Riemann invariants is%
\begin{equation*}
r_{t}=(A^{\prime \prime }(s)+B^{\prime \prime }(r)+r)r_{x}\text{, \ \ \ }%
s_{t}=(A^{\prime \prime }(s)+B^{\prime \prime }(r)+s)s_{x}\text{.}
\end{equation*}%
This system is natural $2-$parametric generalization of gas dynamics (\ref%
{gas}) (see below).

The goal of this paper is a complete description of $N-$component
generalization of above-mentioned formulas.

In section 2 of this paper, so-called ''$\varepsilon -$systems'' are
introduced. All their properties like conservation laws and commuting flows
are described. The corresponding hydrodynamic chain is found by the natural
introduction of moments.

In section 3, some properties such transformations between different
representations of this hydrodynamic chain are obtained.

In section 4, all possible hydrodynamic reductions are found. Particular and
important reductions of this hydrodynamic chain are emphasized.

In section 5, a generating functions of conservation law densities,
commuting flows and solutions (by the Tsarev generalized hodograph method)
for these hydrodynamic reductions are constructed.

In section 6, a general solution of these hydrodynamic type systems is
presented.

In section 7, new ($2+1$)\ integrable hydrodynamic type systems are found.

In section 8, another hydrodynamic chain is presented, all its hydrodynamic
reductions are described.

In section 9, we discuss some still open problems: Hamiltonian structures
and integrable dispersive extensions of hydrodynamic chains and their
reductions.

In section 10 (Conclusion), we describe a general situation in theory of
hydrodynamic chains.

\section{''$\protect\varepsilon -$systems''.}

This class of integrable hydrodynamic type systems%
\begin{equation}
r_{t}^{i}=[r^{i}-\varepsilon \overset{N}{\underset{m=1}{\sum }}%
r^{m}]r_{x}^{i}\text{, \ \ \ \ \ \ }i=1\text{, }2\text{, ..., }N,
\label{one}
\end{equation}%
where $\varepsilon $ is an arbitrary constant, was established in \cite%
{Maks3} (also see \cite{Fer+Maks}, \cite{Maks}, \cite{Maks1}, \cite{Maks4}).
These hydrodynamic type systems (\ref{one}) and its commuting flows (see
below) we shall call ''$\varepsilon -$systems''. The particular case $N=2$
plays important role in gas dynamics (see (\ref{gas}), where the adiabatic
index $\gamma =\frac{3-2\varepsilon }{1-2\varepsilon }$), in field theory ($%
\gamma =1$, Born-Infeld equation), in nonlinear optics ($\gamma =2$, the
dispersionless limit of nonlinear Shrodinger equation, see (\ref{gas})) and
in fluid dynamics ($\gamma =4$, the dispersionless limit of the second
commuting flow to the Boussinesq equation). Also, ''$\varepsilon -$systems''
(for arbitrary $N$) are well known in differential geometry ($\varepsilon
=-1/2$, elliptic coordinates, see, for instance, \cite{Maks1};
dispersionless limit of Coupled KdV, see, for instance, \cite{Fer}), in
soliton theory ($\varepsilon =1$, some particular solutions of
linearly-degenerated systems are multi-gap solutions of KdV, see \cite{Ferap}%
), in biology and chemistry ($\varepsilon =-1$, chromatography,
electrophoresis, isotahophoresis). Moreover, a general solution can be found
explicitly (see \cite{Maks}), for instance, in one-atomic ($\gamma =\frac{5}{%
3}$, $\varepsilon =-1$), two-atomic ($\gamma =\frac{7}{5}$, $\varepsilon =-2$%
) gases (see (\ref{one})) and their generalization for arbitrary $N$ and
arbitrary \textbf{integer} $\varepsilon $. Thus, obvious aim is to extend a
class of integrable hydrodynamic type systems starting from (\ref{one}) with
preservation of some properties.

The hydrodynamic type system (\ref{one}) has a generation function $\mu $ of
conservation law densities; when $\lambda \rightarrow \infty $%
\begin{equation}
\mu \equiv \overset{N}{\underset{m=1}{\prod }}(1-r_{m}/\lambda
)^{-\varepsilon }=1+a_{1}/\lambda +a_{2}/\lambda ^{2}+...,  \label{six}
\end{equation}%
when $\lambda \rightarrow 0$ (up to constant multiplier)%
\begin{equation}
\mu \equiv \overset{N}{\underset{m=1}{\prod }}(r_{m}-\lambda )^{-\varepsilon
}=b_{0}+b_{1}\lambda +b_{2}\lambda ^{2}+...  \label{new}
\end{equation}%
The first series (\ref{six}) is a series of polynomial conservation law
densities $a_{k}$ with respect to Riemann invariants (this is analogue of
Kruskal series of conservation law densities for integrable dispersive
systems like Korteweg de Vries equation). We shall call them as ''higher''
(or ''positive'') conservation law densities correspondingly their
homogeneity. Coefficients $b_{k}$ we shall call as ''lower'' (or
''negative'') conservation law densities (they play role as a ''new''
conservation law densities appearing under Miura type transformation in
theory of integrable dispersive systems).

It is easy to check, that any commuting flow (so, every Riemann invariant $%
r^{i}$ is a function of three independent variables $x$, $t$, $\tau $)%
\begin{equation}
r_{\tau }^{i}=w_{(\varepsilon )}^{i}(\mathbf{r})r_{x}^{i}  \label{coma}
\end{equation}%
to hydrodynamic type system (\ref{one}) has velocities%
\begin{equation}
w_{(\varepsilon )}^{i}(\mathbf{r})=\partial _{i}h_{(-\varepsilon )},
\label{pum}
\end{equation}%
where $h_{(-\varepsilon )}$ is some conservation law density of ''($%
-\varepsilon $)$-$system''%
\begin{equation*}
r_{t}^{i}=[r^{i}+\varepsilon \overset{N}{\underset{m=1}{\sum }}%
r^{m}]r_{x}^{i}\text{, \ \ \ \ \ \ }i=1\text{, }2\text{, ..., }N.
\end{equation*}%
Since, ''$\varepsilon -$systems'' and ''($-\varepsilon $)-systems'' have
generating functions of conservation law densities such that%
\begin{equation}
\mu _{(\varepsilon )}\cdot \mu _{(-\varepsilon )}=1,  \label{inv}
\end{equation}%
then a generating function of commuting flows to (\ref{one}) in Riemann
invariants is (see (\ref{coma}) and (\ref{pum}))%
\begin{equation}
r_{\tau (\tilde{\lambda})}^{i}=\frac{1}{(1-r^{i}/\tilde{\lambda})\tilde{\mu}}%
r_{x}^{i},  \label{cym}
\end{equation}%
and in conservative form (sf. \ref{6}) is%
\begin{equation}
\mu _{\tau (\tilde{\lambda})}=\frac{\tilde{\lambda}}{\tilde{\lambda}-\lambda
}\partial _{x}(\frac{\mu }{\tilde{\mu}}),  \label{zero}
\end{equation}%
where $\tilde{\mu}\equiv \mu (\tilde{\lambda})$ (see (\ref{six}) and (\ref%
{new})).

\textit{Higher} commuting flows can be obtained from (see (\ref{six}))%
\begin{equation*}
\tilde{\mu}\equiv \overset{N}{\underset{m=1}{\prod }}(1-r_{m}/\tilde{\lambda}%
)^{-\varepsilon }=1+a_{1}/\tilde{\lambda}+a_{2}/\tilde{\lambda}^{2}+...
\end{equation*}%
and formal series%
\begin{equation*}
\partial _{\tau (\tilde{\lambda})}=\partial _{t_{0}}+\frac{1}{\tilde{\lambda}%
}\partial _{t_{1}}+\frac{1}{\tilde{\lambda}^{2}}\partial _{t_{1}}+...
\end{equation*}%
when $\tilde{\lambda}\rightarrow \infty $. The corresponding generating
functions of conservation laws are%
\begin{equation}
\partial _{t_{k}}\mu =\partial _{x}[\mu \underset{m=0}{\overset{k}{\sum }}%
\tilde{a}_{m}\lambda ^{k-m}],\ \ \ \ \ k=0,1,2,...  \label{ra}
\end{equation}%
where%
\begin{equation*}
\tilde{a}_{0}=a_{0}=1\text{, \ \ \ \ \ }\tilde{a}_{1}=-a_{1}\text{, \ \ \ \
\ }\tilde{a}_{n}=-a_{n}-\underset{m=1}{\overset{n-1}{\sum }}\tilde{a}%
_{m}a_{n-k}\text{, \ \ \ \ }n=2\text{, }3\text{, ...}
\end{equation*}%
The corresponding \textit{higher} commuting flows (in Riemann invariants) are%
\begin{equation}
\partial _{t_{k}}r^{i}=[\underset{m=0}{\overset{k}{\sum }}\tilde{a}%
_{m}(r^{i})^{k-m}]r_{x}^{i},\ \ \ \ \ k=0,1,2,...  \label{ma}
\end{equation}%
The \textit{lower} commuting flows can be obtained from (see (\ref{new}))%
\begin{equation*}
\tilde{\mu}\equiv \overset{N}{\underset{m=1}{\prod }}(r_{m}-\tilde{\lambda}%
)^{-\varepsilon }=b_{0}+b_{1}\tilde{\lambda}+b_{2}\tilde{\lambda}^{2}+...
\end{equation*}%
and formal series%
\begin{equation*}
\partial _{\tau (\tilde{\lambda})}=\tilde{\lambda}\partial _{t_{-1}}+\tilde{%
\lambda}^{2}\partial _{t_{-2}}+\tilde{\lambda}^{3}\partial _{t_{-3}}+...,
\end{equation*}%
when $\tilde{\lambda}\rightarrow 0$. The corresponding generating functions
of conservation laws are%
\begin{equation}
\partial _{t_{-k-1}}\mu =\partial _{x}[\mu \underset{m=0}{\overset{k}{\sum }}%
\tilde{b}_{m}\lambda ^{m-k-1}]\text{, \ \ \ \ \ }k=0\text{, }1\text{, }2%
\text{, ... ,}  \label{pa}
\end{equation}%
where%
\begin{equation*}
\tilde{b}_{0}=\frac{1}{b_{0}}\text{, \ \ \ \ }\tilde{b}_{k}=-\frac{1}{b_{0}}%
\underset{m=0}{\overset{k-1}{\sum }}\tilde{b}_{m}b_{k-m}\text{, \ \ \ \ \ \ }%
k=1\text{, }2\text{, ...}
\end{equation*}%
The corresponding \textit{lower} commuting flows (in Riemann invariants) are%
\begin{equation}
\partial _{t_{-k-1}}r^{i}=[\underset{m=0}{\overset{k}{\sum }}\tilde{b}%
_{m}(r^{i})^{m-k-1}]r_{x}^{i}\text{, \ \ \ \ \ }k=0\text{, }1\text{, }2\text{%
, ... ,}  \label{ga}
\end{equation}%
If $\lambda \rightarrow \infty $, $\tilde{\lambda}\rightarrow \infty $, all
the \textit{higher} conservation laws for the \textit{higher} commuting
flows are%
\begin{equation}
\partial _{t_{k}}a_{m}=\partial _{x}[\underset{s=0}{\overset{k}{\sum }}%
\tilde{a}_{s}a_{k+m-s}]\text{, \ \ \ \ \ }k=0\text{, }1\text{, }2\text{, ...}
\label{firs}
\end{equation}%
If $\lambda \rightarrow 0$, $\tilde{\lambda}\rightarrow \infty $, all the
\textit{lower} conservation laws for the \textit{higher} commuting flows are%
\begin{equation}
\partial _{t_{n}}b_{k}=\partial _{x}[\underset{s=0}{\overset{k}{\sum }}b_{s}%
\tilde{a}_{n+s-k}]\text{, \ \ \ }k\leqslant n\text{, \ \ \ \ \ }\partial
_{t_{n}}b_{k}=\partial _{x}[\underset{s=0}{\overset{n}{\sum }}\tilde{a}%
_{s}b_{k+s-n}]\text{, \ \ \ }k\geqslant n.  \label{9}
\end{equation}%
If $\lambda \rightarrow 0$, $\tilde{\lambda}\rightarrow 0$, all the \textit{%
lower} conservation laws for \textit{lower} commuting flows are%
\begin{equation}
\partial _{t_{-n-1}}b_{m}=\partial _{x}[\underset{k=0}{\overset{n}{\sum }}%
\tilde{b}_{k}b_{n+m+1-k}]\text{, \ \ \ \ \ }n=0\text{, }1\text{, }2\text{,
...}  \label{za}
\end{equation}%
If $\lambda \rightarrow \infty $, $\tilde{\lambda}\rightarrow 0$, all the
\textit{higher} conservation laws for the \textit{lower} commuting flows are%
\begin{equation}
\partial _{t_{-n-1}}a_{m+1}=\partial _{x}[\underset{k=0}{\overset{m}{\sum }}%
a_{s}\tilde{b}_{n+s-m}]\text{,\ }m\leqslant n\text{,\ \ }\partial
_{t_{-n-1}}a_{m+1}=\partial _{x}[\underset{k=0}{\overset{n}{\sum }}\tilde{b}%
_{s}a_{m+s-n}]\text{,\ }m\geqslant n\text{.}  \label{8}
\end{equation}%
All these above formulas can be easily checked by a direct calculations.

For instance, the initial system (\ref{one}) has the generating function of
conservation laws%
\begin{equation}
\partial _{t}\mu =\partial _{x}[(\lambda -a_{1})\mu ],  \label{five}
\end{equation}%
where an infinite set of the \textit{positive} (polynomial) conservation
laws is%
\begin{equation}
\partial _{t_{1}}a_{k}=\partial _{x}[a_{k+1}-a_{1}a_{k}]\text{, \ \ \ \ }k=1%
\text{, }2\text{, ...}  \label{four}
\end{equation}%
and an infinite set of the \textit{negative} conservation laws is%
\begin{equation}
\partial _{t_{1}}b_{0}=\partial _{x}(-a_{1}b_{0})\text{, \ \ \ \ \ }\partial
_{t_{1}}b_{k}=\partial _{x}[b_{k-1}-a_{1}b_{k}]\text{, \ \ \ \ }k=1\text{, }2%
\text{, ...}  \label{ba}
\end{equation}%
The second commuting flow (see (\ref{one}) and (\ref{ma}))%
\begin{equation*}
r_{t_{2}}^{i}=[(r^{i})^{2}-\varepsilon r^{i}\underset{m=0}{\overset{N}{\sum }%
}r^{m}+\frac{\varepsilon ^{2}}{2}(\underset{m=0}{\overset{N}{\sum }}%
r^{m})^{2}-\frac{\varepsilon }{2}\underset{m=0}{\overset{N}{\sum }}%
(r^{m})^{2}]r_{x}^{i}\text{, \ \ \ }i=1\text{, }2\text{, ... , }N
\end{equation*}%
has the generating function of conservation laws (see (\ref{ra}))%
\begin{equation*}
\mu _{t_{2}}=\partial _{x}[(\lambda ^{2}-a_{1}\lambda +a_{1}^{2}-a_{2})\mu ],
\end{equation*}%
where an infinite set of the \textit{positive} (polynomial) conservation
laws is (see (\ref{firs}))%
\begin{equation}
\partial _{t_{2}}a_{k}=\partial
_{x}[a_{k+2}-a_{1}a_{k+1}+(a_{1}^{2}-a_{2})a_{k}].  \label{second}
\end{equation}%
an infinite set of the \textit{negative} conservation laws is (see (\ref{9}))%
\begin{eqnarray*}
\partial _{t_{2}}b_{0} &=&\partial _{x}[b_{0}(a_{1}^{2}-a_{2})]\text{, \ \ \
\ \ }\partial _{t_{2}}b_{1}=\partial _{x}[b_{1}(a_{1}^{2}-a_{2})-a_{1}b_{0}]%
\text{,} \\
\partial _{t_{2}}b_{k} &=&\partial
_{x}[b_{k}(a_{1}^{2}-a_{2})-a_{1}b_{k-1}+b_{k-2}]\text{, \ \ \ \ }k=2\text{,
}3\text{, ...}
\end{eqnarray*}%
The first \textit{negative} ($\tilde{\lambda}\rightarrow 0$) commuting flow
is (see (\ref{ga}))%
\begin{equation}
r_{t_{-1}}^{i}=\frac{1}{b_{0}r^{i}}r_{x}^{i}=\frac{\overset{N}{\underset{m=1}%
{\prod }}(r^{m})^{\varepsilon }}{r^{i}}r_{x}^{i},  \label{com}
\end{equation}%
where the generating function of conservation laws is (see (\ref{pa}))%
\begin{equation}
\mu _{t_{-1}}=\partial _{x}\frac{\mu }{\lambda b_{0}},  \label{next}
\end{equation}%
where an infinite set of \textit{negative} conservation laws is (see (\ref%
{za}))%
\begin{equation}
\partial _{t_{-1}}b_{k}=\partial _{x}\frac{b_{k+1}}{b_{0}}\text{, \ \ \ \ \ }%
k=0\text{, }1\text{, }2\text{, ...}  \label{min}
\end{equation}%
and an infinite set of \textit{positive} conservation laws is (see (\ref{8}))%
\begin{equation}
\partial _{t_{-1}}a_{1}=\partial _{x}\frac{1}{b_{0}}\text{, \ \ \ \ \ \ }%
\partial _{t_{-1}}a_{k+1}=\partial _{x}\frac{a_{k}}{b_{0}}\text{, \ \ \ \ \ }%
k=1\text{, }2\text{, ...}  \label{ab}
\end{equation}

\begin{remark}
Reciprocal transformation (see the first equation in (\ref{ba}))%
\begin{equation*}
dy_{-1}=b_{0}dx-a_{1}b_{0}dt_{1}\text{, \ \ \ \ \ }dz=dt_{1},
\end{equation*}%
connects system (\ref{one}) (also, see (\ref{four}) and (\ref{ba})) and its
commuting flow (\ref{com}) (also, see (\ref{min}) and (\ref{ab})).

Reciprocal transformation (see (\ref{five}))%
\begin{equation*}
dy=\mu \lbrack dx+(\lambda -a_{1})dt_{1}]\text{, \ \ \ \ }dz=dt_{1}
\end{equation*}%
connects system (\ref{one}) and generating function of its commuting flows
(see (\ref{coma})-(\ref{zero}))%
\begin{equation*}
\partial _{t}a_{1}=-\lambda \partial _{x}(1/\mu ).
\end{equation*}
\end{remark}

\section{New hydrodynamic chain}

The hydrodynamic type system (\ref{one}) can be rewritten as an infinite
momentum chain%
\begin{equation}
\partial _{t}c_{k}=\partial _{x}c_{k+1}-c_{1}\partial _{x}c_{k}\text{, \ \ \
\ }k=0\text{, }\pm 1\text{, }\pm 2\text{, ...}  \label{tri}
\end{equation}%
where $N$ first moments $c_{k}$ ($k=1$, $2$, ..., $N$) are functionally
independent%
\begin{equation}
c_{0}=\varepsilon \overset{N}{\underset{m=1}{\sum }}\ln r^{m}\text{, \ \ \ \
\ }c_{k}=\frac{\varepsilon }{k}\overset{N}{\underset{m=1}{\sum }}%
(r^{m})^{k},\ \ \text{\ \ \ }k=\pm 1\text{, }\pm 2\text{, ...}  \label{Rimm}
\end{equation}%
Thus, all $a_{k}$, $b_{k}$ and $c_{k}$ can be expressed via each other (see
below). This is invertible transformation.

However, now we can start our investigation namely from hydrodynamic chain
written in form (\ref{tri}) or, for instance, (\ref{four}) without reference
on original hydrodynamic type system (\ref{one}) (and explicit expressions (%
\ref{Rimm})). If now we restrict our infinite momentum chain to $N-$%
component case, then just one particular solution obviously is the
hydrodynamic type system (\ref{one}) (also, see (\ref{Rimm})). How to find
all other possible reductions? The answer will be done in next section.

\begin{remark}
For the first time, this hydrodynamic chain (\ref{four}) has been derived
(in another terms) by S.J. Alber (see \cite{Alber}; also it has been
independently obtained in another context by V.G. Mikhalev, see \cite{Mikh}%
), and recently by L.M. Alonso and A.B. Shabat (see \cite{Shabat}; they
describe \textbf{mostly} differential reductions and \textbf{very particular}
hydrodynamic reductions, in our article we describe \textbf{all possible}
hydrodynamic reductions). Actually, the starting point of their
investigations was the hydrodynamic type ''$\varepsilon -$system'', when $%
\varepsilon =-1/2$, related by generalized reciprocal transformation with
averaged (by the Whitham method) integrable systems (determined by scalar
second order spectral transform with energy-dependent potential, see also
mentioned references) related with hyperelliptic surfaces.
\end{remark}

The generating function of conservation laws for the hydrodynamic chain (\ref%
{tri}) is exactly (\ref{five}), where (sf. (\ref{six}))%
\begin{equation}
\mu =1+\overset{\infty }{\underset{k=1}{\sum }}a_{k}\lambda ^{-k}=\exp [%
\overset{\infty }{\underset{k=1}{\sum }}c_{k}\lambda ^{-k}]\text{, \ \ \ \ \
}c_{1}\equiv a_{1}  \label{nul}
\end{equation}%
and $\lambda \rightarrow \infty $. The hydrodynamic chain (\ref{ba})
satisfies the same generating function (\ref{five}), where (sf. (\ref{new}))%
\begin{equation}
\mu =\overset{\infty }{\underset{k=0}{\sum }}b_{k}\lambda ^{k}=\exp [-%
\overset{\infty }{\underset{k=0}{\sum }}c_{-k}\lambda ^{k}]\text{, \ \ \ \ \
\ }c_{0}\equiv -\ln b_{0}  \label{null}
\end{equation}%
and $\lambda \rightarrow 0$.

All the \textit{positive} commuting flows in field variables $c_{k}$ are%
\begin{equation*}
\partial _{t_{n}}c_{k}=\overset{n}{\underset{m=0}{\sum }}\tilde{a}%
_{m}c_{k+n-m,x}\text{, \ \ \ \ \ \ }k=0\text{, }\pm 1\text{, }\pm 2\text{,
...}
\end{equation*}%
where the generating function of conservation law densities for arbitrary
\textit{positive} commuting flow is (\ref{ra}); all \textit{negative} flows
are%
\begin{equation*}
\partial _{t_{-n-1}}c_{k}=\overset{n}{\underset{m=0}{\sum }}\tilde{b}%
_{m}c_{k+m-n,x}\text{, \ \ \ \ \ \ }k=0\text{, }\pm 1\text{, }\pm 2\text{,
... ,}
\end{equation*}%
where the generating function of conservation law densities for arbitrary
\textit{negative} commuting flow is (\ref{pa}). All these \textit{negative}
flows can be obtained from \textit{positive} commuting flows (see above) by
the reciprocal transformation (see \textbf{Remark} of previous section). For
instance, the first \textit{negative} flow is%
\begin{equation}
\partial _{t_{-1}}c_{k}=e^{c_{0}}\partial _{x}c_{k-1}\text{, \ \ \ \ \ \ }k=0%
\text{, }\pm 1\text{, }\pm 2\text{, ...}  \label{minus}
\end{equation}

\begin{remark}
Obviously, values $\tilde{a}_{k}$ and $\tilde{b}_{k}$ can be expressed from
analogues of (\ref{nul}) and (\ref{null}) (see (\ref{zero}))%
\begin{eqnarray*}
\frac{1}{\mu } &=&1+\overset{\infty }{\underset{k=1}{\sum }}\tilde{a}%
_{k}\lambda ^{-k}=\exp [-\overset{\infty }{\underset{k=1}{\sum }}%
c_{k}\lambda ^{-k}]\text{, \ \ \ \ \ }\tilde{a}_{1}\equiv -c_{1}\text{, \ \ }%
\lambda \rightarrow \infty . \\
&& \\
\frac{1}{\mu } &=&\overset{\infty }{\underset{k=0}{\sum }}\tilde{b}%
_{k}\lambda ^{k}=\exp [\overset{\infty }{\underset{k=0}{\sum }}c_{-k}\lambda
^{k}]\text{, \ \ \ \ \ \ }\tilde{b}_{0}\equiv e^{c_{0}}\text{, \ \ \ \ \ }%
\lambda \rightarrow 0.
\end{eqnarray*}
\end{remark}

Thus, hydrodynamic chain (\ref{tri}) can be expressed via different moments (%
$a_{k}$, $b_{k}$), see, for instance Remark in section 6.

Relationship (\ref{nul}) \textit{positive} moments $c_{k}$ and \textit{%
positive} conservation law densities $a_{k}$ can be expressed explicitly by
next four recursive formulas, where the first of them%
\begin{equation*}
da_{k+1}=\underset{m=1}{\overset{k}{\sum }}a_{m}dc_{k+1-m}+dc_{k+1}\text{, \
\ \ \ \ \ }k=0,1,2,...
\end{equation*}%
is consequence of three local symmetry operators acting on space of
conservation law densities $a_{k}$: the shift operator%
\begin{equation*}
\hat{\delta}a_{k+1}=\frac{\partial a_{k+1}}{\partial c_{1}}=a_{k}\text{, \ \
\ \ \ \ \ \ \ }k=0,1,2,...,
\end{equation*}%
the scaling operator%
\begin{equation*}
\hat{R}a_{k}=\underset{m=1}{\overset{\infty }{\sum }}mc_{m}\frac{\partial
a_{k}}{\partial c_{m}}=ka_{k}\text{, \ \ \ \ \ \ \ \ \ }k=0,1,2,...
\end{equation*}%
and projective operator%
\begin{equation*}
\hat{S}a_{k}=[c_{1}+\underset{m=1}{\overset{\infty }{\sum }}(m+1)c_{m+1}%
\frac{\partial }{\partial c_{m}}]a_{k}=(k+1)a_{k+1}\text{, \ \ \ \ \ \ \ \ \
}k=0,1,2,...
\end{equation*}

\section{Finite-component reductions}

\begin{theorem}
The hydrodynamic type system (\ref{red}) with an arbitrary number of
components is embedded into the hydrodynamic chain (\ref{tri}) if and only if%
\begin{equation}
\text{ \ \ \ \ }V_{i}=f_{i}(r^{i})-c_{1}\text{, \ \ \ \ \ \ }c_{1}=\underset{%
m=1}{\overset{N}{\sum }}\psi _{m}(r^{m})\text{,}  \label{blue}
\end{equation}%
where $f_{i}(r^{i})$ and $\psi _{k}(r^{k})$ are arbitrary functions.
\end{theorem}

\begin{proof}
Any reductions are compatible with given hydrodynamic chain (\ref{tri}) if
every $c_{k}$ can be expressed as a function of just $N$ independent Riemann
invariants $r^{i}$. Then one obtains%
\begin{equation*}
V_{i}\partial _{i}c_{k}=\partial _{i}c_{k+1}-c_{1}\partial _{i}c_{k}\text{,
\ \ \ \ \ }i=1\text{, }2\text{, ..., }N\text{,\ \ \ \ \ \ \ }k=0\text{, }\pm
1\text{, }\pm 2\text{, ...}
\end{equation*}%
It is easy to see, that%
\begin{equation*}
\partial _{i}c_{k+1}=(V_{i}+c_{1})^{k}\partial _{i}c_{1}\text{, \ \ \ \ \ }%
i=1\text{, }2\text{, ..., }N\text{,\ \ \ \ \ \ \ }k=0\text{, }\pm 1\text{, }%
\pm 2\text{, ...}
\end{equation*}%
Thus, the second derivatives%
\begin{equation*}
\partial _{j}[(V_{i}+c_{1})^{k}\partial _{i}c_{1}]=\partial
_{i}[(V_{j}+c_{1})^{k}\partial _{j}c_{1}]\text{, \ \ \ }i=1\text{, }2\text{,
..., }N\text{,\ \ \ \ \ \ \ }k=0\text{, }\pm 1\text{, }\pm 2\text{, ...}
\end{equation*}%
yield the general reduction (\ref{blue}).
\end{proof}

Thus, all moments are%
\begin{equation*}
c_{k}=\underset{m=1}{\overset{N}{\sum }}\overset{r^{m}}{\int }[f_{m}(\lambda
)]^{k-1}d\psi _{m}(\lambda )\text{, \ \ \ \ \ }k=0\text{, }\pm 1\text{, }\pm
2\text{, ...}
\end{equation*}

\begin{remark}
The hydrodynamic type systems%
\begin{equation}
r_{t_{-1}}^{i}=W^{i}(\mathbf{r})r_{x}^{i}\text{, \ \ \ \ }i=1\text{, }2\text{%
, ... , }N  \label{minna}
\end{equation}%
embedded into first negative flow (\ref{minus}) (see (\ref{tri})) can be
found in the same way (see (\ref{blue})): from $W^{i}\partial
_{i}c_{k}=e^{c_{0}}\partial _{i}c_{k-1}$ one can obtain $\partial _{i}c_{k}=(%
\frac{e^{c_{0}}}{W^{i}})^{k}\partial _{i}c_{0}$. In comparison with $%
\partial _{i}c_{k}=(V_{i}+c_{1})^{k}\partial _{i}c_{0}$ one can obtain%
\begin{equation}
W^{i}=\frac{1}{f_{i}(r^{i})}\exp [\underset{m=1}{\overset{N}{\sum }}\overset{%
r^{m}}{\int }\frac{d\psi _{m}(\lambda )}{f_{m}(\lambda )}].  \label{pin}
\end{equation}
\end{remark}

Next task is how to write hydrodynamic reductions in \textit{closed} form
via special variables like conservation law densities (see, e.g. (\ref{four}%
)). It means, that all higher moments $a_{N+k}$ must be expressed via lower
moments $a_{k}$ ($k=1$, $2$, ... , $N$). In particular case (\ref{one}), all
higher moments $a_{N+k}$ are polynoms, which can be found from relations
(see (\ref{six}))%
\begin{equation*}
0=[(1+\lambda a_{1}+\lambda ^{2}a_{2}+...+\lambda ^{N}a_{N}+\lambda
^{N+1}a_{N+1}+...)^{-1/\varepsilon }]^{(N+k)}\text{, \ \ \ }k=1\text{, }2%
\text{, ...}
\end{equation*}
For example, the first higher moment $a_{N+1}$ can be found from more
compact recursive relation%
\begin{equation*}
\overset{N+1}{\underset{k=1}{\sum }}\frac{\Gamma (1-1/\varepsilon
)z_{k}^{(N+1)}}{\Gamma (1-k-1/\varepsilon )\Gamma (k+1)}=0,
\end{equation*}%
where $z_{k}^{(N+1)}$ is coefficient of series%
\begin{equation*}
\lbrack a_{1}+a_{2}\lambda +a_{3}\lambda ^{2}+...+a_{N+2-k}\lambda
^{N+1-k}]^{k}=a_{1}^{k}+\lambda ka_{1}^{k-1}a_{2}+...+\lambda
^{N+1-k}z_{k}^{(N+1)}+...
\end{equation*}%
All higher moments $a_{N+k}$ are polynoms of lower moments $a_{k}$, e.g.:

\begin{equation*}
\begin{array}{c}
a_{3}=\frac{1+\varepsilon }{\varepsilon }a_{1}a_{2}-\frac{(1+\varepsilon
)(1+2\varepsilon )}{6\varepsilon ^{2}}a_{1}^{3}\text{, \ \ \ \ \ }N=2, \\
\\
a_{4}=\frac{1+\varepsilon }{2\varepsilon }(2a_{1}a_{3}+a_{2}^{2})-\frac{%
(1+\varepsilon )(1+2\varepsilon )}{2\varepsilon ^{2}}a_{1}^{2}a_{2}+\frac{%
(1+\varepsilon )(1+2\varepsilon )(1+3\varepsilon )}{24\varepsilon ^{3}}%
a_{1}^{4}\text{, \ \ \ \ }N=3, \\
\\
a_{5}=\frac{1+\varepsilon }{\varepsilon }(a_{1}a_{4}+a_{2}a_{3})-\frac{%
(1+\varepsilon )(1+2\varepsilon )}{2\varepsilon ^{2}}%
(a_{1}^{2}a_{3}+a_{1}a_{2}^{2})+ \\
\\
\frac{(1+\varepsilon )(1+2\varepsilon )(1+3\varepsilon )}{6\varepsilon ^{3}}%
a_{1}^{3}a_{2}-\frac{(1+\varepsilon )(1+2\varepsilon )(1+3\varepsilon
)(1+4\varepsilon )}{120\varepsilon ^{4}}a_{1}^{5}\text{, \ \ \ \ \ \ \ }N=4%
\text{,}%
\end{array}%
\end{equation*}%
and so on... First exceptional case, is chromatography phenomena ($%
\varepsilon =-1)$, then

\begin{equation*}
\begin{array}{c}
a_{3}=-a_{1}a_{2}+\frac{1}{6}a_{1}^{3}\text{, \ \ \ \ \ }N=2, \\
\\
a_{4}=-\frac{1}{2}(2a_{1}a_{3}+a_{2}^{2})+\frac{1}{2}a_{1}^{2}a_{2}-\frac{1}{%
12}a_{1}^{4}\text{, \ \ \ \ }N=3, \\
\\
a_{5}=-(a_{1}a_{4}+a_{2}a_{3})+\frac{1}{2}(a_{1}^{2}a_{3}+a_{1}a_{2}^{2})-%
\frac{1}{3}a_{1}^{3}a_{2}+\frac{1}{20}a_{1}^{5}\text{, \ \ \ \ \ \ \ }N=4%
\text{;}%
\end{array}%
\end{equation*}%
the second exceptional case is (dispersionless limits of Coupled KdV and
Coupled Harry Dym) $\varepsilon =-1/2$, then first higher moment $a_{k+1}$
is just quadratic expression via lower moments $a_{k}$. It means, that
corresponding hydrodynamic type systems have at least one local Hamiltonian
structure (see \cite{Maks+Tsar}); actually, the most number of local
Hamiltonian structures is ($N+1$), \textit{iff} $\varepsilon =-1/2$, see %
\cite{Fer})). The third exceptional case is $\varepsilon =-1/M$, where $M=3$%
, $4$, ... Then all higher moments will be quickly trancated (see above).

\section{Commuting flows and reductions}

The corresponding linear system for conservation law densities (see \cite%
{Tsar}) is%
\begin{equation}
\partial _{ik}h=\frac{\psi _{k}^{\prime }(r^{k})}{f_{i}(r^{i})-f_{k}(r^{k})}%
\partial _{i}h-\frac{\psi _{i}^{\prime }(r^{i})}{f_{i}(r^{i})-f_{k}(r^{k})}%
\partial _{k}h\text{, \ \ \ \ \ \ }i\neq k.  \label{third}
\end{equation}%
The general solution of such system is determined by $N$ functions of a
single variable (also, see \cite{Tsar}). In a particular, but very
important, case of ''$\varepsilon -$systems'' (see, for example, (\ref{one}%
)) this linear system (\ref{third}) is exactly $N-$component generalization
of the Euler-Darboux-Poisson system (see \cite{Maks})%
\begin{equation}
\partial _{ik}h=\frac{\varepsilon }{r^{i}-r^{k}}[\partial _{i}h-\partial
_{k}h]\text{, \ \ \ \ \ \ }i\neq k.  \label{Pois}
\end{equation}%
At first, it is necessary to find a generating function of conservation law
densities $\mu $, which can be found in comparison with (\ref{five}) and (%
\ref{blue})%
\begin{equation}
\mu (\mathbf{r},\lambda )=\exp \underset{k=1}{\overset{N}{\sum }}\overset{%
r^{k}}{\int }\frac{d\psi _{k}(\tilde{\lambda})}{\lambda -f_{k}(\tilde{\lambda%
})}.  \label{first}
\end{equation}%
This formula (\ref{first}) simplifies in case of ''$\varepsilon -$systems''
(see (\ref{six}) and (\ref{new}), that was well known in case $N=2$, for
example, see \cite{Darboux}). Velocities $w^{i}$ of the commuting flows
(i.e. Riemann invariants $r^{i}$ are considered as the functions
simultaneously of three independent variables $x$, $t$, $\tau $)%
\begin{equation}
r_{\tau }^{i}=w^{i}(\mathbf{r})r_{x}^{i}\text{, \ \ \ \ \ \ }i=1\text{, }2%
\text{, ..., }N  \label{fifth}
\end{equation}%
can be found as the solutions of another linear system (also see \cite{Tsar})%
\begin{equation}
\partial _{i}w^{k}=-\frac{\psi _{i}^{\prime }}{f_{i}(r^{i})-f_{k}(r^{k})}%
(w^{i}-w^{k})\text{, \ \ \ \ \ \ }i\neq k.  \label{sixth}
\end{equation}

\begin{theorem}
Any solutions of the linear system (\ref{sixth}) are connected with the
solutions of another linear system (sf. (\ref{third}))%
\begin{equation*}
\partial _{ik}\tilde{h}=-\frac{\psi _{k}^{\prime }(r^{k})}{%
f_{i}(r^{i})-f_{k}(r^{k})}\partial _{i}\tilde{h}+\frac{\psi _{i}^{\prime
}(r^{i})}{f_{i}(r^{i})-f_{k}(r^{k})}\partial _{k}\tilde{h}\text{, \ \ \ \ \
\ }i\neq k
\end{equation*}%
by the differential substitution of the first order%
\begin{equation*}
w^{i}=\frac{1}{\psi _{i}^{\prime }}\partial _{i}\tilde{h}.
\end{equation*}
\end{theorem}

Thus, the generating function of solutions (for the hydrodynamic type
systems (\ref{red}) and (\ref{blue}), (\ref{minna}) and (\ref{pin}); cf. (%
\ref{one}) and (\ref{com})) by Tsarev generalized hodograph method (also see %
\cite{Tsar}) is%
\begin{equation}
\begin{array}{c}
x+[f_{i}(r^{i})-\underset{k=1}{\overset{N}{\sum }}\psi _{k}(r^{k})]t_{1}+%
\frac{1}{f_{i}(r^{i})}\exp [\underset{m=1}{\overset{N}{\sum }}\overset{r^{m}}%
{\int }\frac{d\psi _{m}(\lambda )}{f_{m}(\lambda )}]t_{-1}= \\
\\
-\frac{1}{\lambda -f_{i}(r^{i})}\exp \left[ -\underset{k=1}{\overset{N}{\sum
}}\overset{r^{k}}{\int }\frac{d\psi _{k}(\tilde{\lambda})}{\lambda -f_{k}(%
\tilde{\lambda})}\right] ,%
\end{array}
\label{hod}
\end{equation}%
where the generating function of commuting flows (see (\ref{cym})) in
Riemann invariants is%
\begin{equation*}
r_{\tau (\tilde{\lambda})}^{i}=\frac{1}{(1-f_{i}(r^{i})/\tilde{\lambda})%
\tilde{\mu}}r_{x}^{i}.
\end{equation*}

\begin{remark}
If one selects monoms%
\begin{equation*}
\psi _{i}(r^{i})=\varepsilon _{i}r^{i},
\end{equation*}%
where $\varepsilon _{k}$ are arbitrary constants, then the generalized ''$%
\varepsilon -$systems''%
\begin{equation}
r_{t}^{i}=[r^{i}-\underset{m=1}{\overset{N}{\sum }}\varepsilon
_{m}r^{m}]r_{x}^{i}.  \label{ten}
\end{equation}%
has the generating function of conservation law densities (cf. (\ref{six})
and (\ref{first}))%
\begin{equation*}
\mu =\underset{m=1}{\overset{N}{\prod }}(1-r_{m}/\lambda )^{-\varepsilon
_{m}}.
\end{equation*}%
In general case ($N$ is arbitrary) the hydrodynamic type system (\ref{ten})
is the natural $N-$parametric reduction of the hydrodynamic chain (\ref{tri}%
). When $N=2$ this system (\ref{ten}) is natural two-parametric
generalization of gas dynamics (\ref{gas}). If we choose (cf. (\ref{Rimm}))%
\begin{equation*}
c_{0}=\overset{N}{\underset{m=1}{\sum }}\varepsilon _{m}\ln r^{m}\text{, \ \
\ \ }c_{k}=\frac{1}{k}\underset{m=1}{\overset{N}{\sum }}\varepsilon
_{m}(r^{m})^{k},\ \ \text{\ \ \ }k=\pm 1\text{, }\pm 2\text{, ...}
\end{equation*}%
then the hydrodynamic type system (\ref{ten}) satisfies for hydrodynamic
chain (\ref{tri}).
\end{remark}

\begin{remark}
In the particular case $f_{i}(r^{i})=\varepsilon _{i}$ ($\varepsilon _{i}$
are arbitrary constants and $\varepsilon _{i}\neq \varepsilon _{k}$ for\ $%
i\neq k$; this is $N-$component generalization of gas dynamics, when the
adiabatic index $\gamma =1$) hydrodynamic type system (\ref{blue})%
\begin{equation*}
r_{t}^{i}=[\varepsilon _{i}-\underset{m=1}{\overset{N}{\sum }}\psi
_{m}(r^{m})]r_{x}^{i}
\end{equation*}%
is ''trivial'' (also $\psi _{i}(r^{i})\neq \func{const}$). In this case, the
linear system (\ref{third}) has constant coefficients%
\begin{equation*}
\frac{\partial ^{2}h}{\partial R^{i}\partial R^{k}}=\frac{1}{\varepsilon
_{i}-\varepsilon _{k}}[\frac{\partial h}{\partial R^{i}}-\frac{\partial h}{%
\partial R^{k}}]\text{, \ \ \ \ \ }i\neq k,
\end{equation*}%
where $R^{i}=\psi _{i}(r^{i})$.
\end{remark}

\section{General solution}

Description of the general solution for the linear system (\ref{third}) is a
very complicated task. Construction of the general solution has been made
only in case of ''$\varepsilon -$systems'' (see \cite{Maks}), when $N$ is
arbitrary; cases $N=2$ and $N=3$ were completely investigated by G. Darboux,
L.P. Eisenhart and T.H. Gronwall (see \cite{Darboux} and \cite{Eizen}). The
basic idea of how to construct a general solution (parameterized by $N$
functions of a single variable, see \cite{Tsar}) of any over-determined
linear systems like (\ref{third}) was presented in \cite{Tsar} by recursive
application of symmetry operators compatible with such systems. However,
here we establish an alternative approach in the spirit of G. Darboux (see %
\cite{Tsar}; also, see the section concerning elliptic coordinates in \cite%
{Darboux}). Elliptic coordinates $\mu _{\alpha }$ ($\alpha =1$, $2$, ..., $N$%
) appear in the theory of integrable hydrodynamic type systems associated
with hyperelliptic curves, i.e. with ''$\varepsilon -$systems'', where $%
\varepsilon =-1/2$. G. Darboux suggested to introduce special variables $%
r^{k}$ ($k=1$, $2$, ..., $N$) for separation of coordinates in Laplace
equation by the following rule (see (\ref{new}), when $\varepsilon =-1/2$)%
\begin{equation*}
\mu _{\alpha }^{2}=\frac{\underset{k=1}{\overset{N}{\prod }}(\gamma _{\alpha
}-r^{k})}{\underset{\beta \neq \alpha }{\prod }(\gamma _{\alpha }-\gamma
_{\beta })}\text{, \ \ \ \ \ \ }g^{ii}=\frac{\underset{\beta =1}{\overset{N}{%
\prod }}(r^{i}-\gamma _{\beta })}{\underset{k\neq i}{\prod }(r^{i}-r^{k})}%
\text{,}
\end{equation*}%
where $\gamma _{\alpha }$ are arbitrary constants (the denominator $\underset%
{\beta \neq \alpha }{\prod }(\gamma _{\alpha }-\gamma _{\beta })$ in the
first expression is just a constant multiplier, which does not affect the
property to be a conservation law density, also see (\ref{five})) and flat
(not constant) metric $g^{ii}(\mathbf{r})$ determined by%
\begin{equation*}
ds^{2}=\underset{\alpha =1}{\overset{N}{\sum }}(d\mu _{\alpha })^{2}=%
\underset{k=1}{\overset{N}{\sum }}g_{kk}(dr^{k})^{2}.
\end{equation*}%
Thus, elliptic coordinates coincide with the Riemann invariants for ''$%
\varepsilon -$systems'', where $\varepsilon =-1/2$. It is easy to generalize
Darboux coordinates $\mu _{\alpha }$ to arbitrary $\varepsilon $%
\begin{equation*}
(\mu _{\alpha })^{-1/\varepsilon }=\frac{\underset{k=1}{\overset{N}{\prod }}%
(\gamma _{\alpha }-r^{k})}{\underset{\beta \neq \alpha }{\prod }(\gamma
_{\alpha }-\gamma _{\beta })}\text{, \ \ \ \ }\alpha =1\text{, }2\text{, ...
, }N.
\end{equation*}%
In this case, all ''$\varepsilon -$systems'', for instance, (\ref{one}) and (%
\ref{com}) can be written explicitly via $\mu _{\alpha }$ in the
conservative form (see (\ref{four}) and (\ref{min}) in the particular case $%
\varepsilon =-1/2$)
\begin{eqnarray}
\partial _{t}\mu _{\alpha } &=&\partial _{x}\left[ \left( \varepsilon
\underset{\beta =1}{\overset{N}{\sum }}(\mu _{\beta })^{-1/\varepsilon
}+\gamma _{\alpha }-\varepsilon \underset{\beta =1}{\overset{N}{\sum }}%
\gamma _{\beta }\right) \mu _{\alpha }\right] \text{,}  \notag \\
&&  \label{g} \\
\text{ \ \ \ \ \ }\partial _{t_{-1}}\mu _{\alpha } &=&\frac{\underset{\beta
=1}{\overset{N}{\prod }}(\gamma _{\beta })^{\varepsilon }}{\gamma _{\alpha }}%
\partial _{x}\left[ \left( 1-\underset{\beta =1}{\overset{N}{\sum }}\frac{%
(\mu _{\beta })^{-1/\varepsilon }}{\gamma _{\beta }}\right) ^{\varepsilon
}\mu _{\alpha }\right] \text{, \ \ \ \ \ }\alpha =1\text{, }2\text{, ... , }%
N.  \notag
\end{eqnarray}

\begin{remark}
Hydrodynamic type systems (\ref{one}) and (\ref{com}) for another set of
moments $E_{k}=\underset{\beta =1}{\overset{N}{\sum }}(\gamma _{\beta
})^{k}(\mu _{\beta })^{-1/\varepsilon }$ (see (\ref{g})) can be written as
the following hydrodynamic chains%
\begin{eqnarray}
\partial _{t_{1}}E_{k} &=&\partial _{x}E_{k+1}+\varepsilon \lbrack E_{0}-%
\underset{\beta =1}{\overset{N}{\sum }}\gamma _{\beta }]E_{k,x}-E_{k}E_{0,x}%
\text{, \ \ \ \ \ \ }k=0\text{, }\pm 1\text{, }\pm 2\text{, ... ,}  \label{z}
\\
\partial _{t_{-1}}E_{k} &=&\underset{\beta =1}{\overset{N}{\prod }}(\gamma
_{\beta })^{\varepsilon }[(1-E_{-1})^{\varepsilon }\partial
_{x}E_{k-1}+(1-E_{-1})^{\varepsilon -1}E_{k-1}E_{-1,x}],  \label{zm}
\end{eqnarray}%
where%
\begin{equation*}
a_{1}=\varepsilon (\underset{\beta =1}{\overset{N}{\sum }}\gamma _{\beta
}-E_{0})\text{, \ \ \ \ \ }b_{0}=\underset{\beta =1}{\overset{N}{\prod }}%
(\gamma _{\beta })^{-\varepsilon }(1-E_{-1})^{-\varepsilon }.
\end{equation*}
\end{remark}

\begin{remark}
The hydrodynamic chain (\ref{z}) is the same as the hydrodynamic chain (\ref%
{four}), because these two chains are connected by invertible transformation
(see (\ref{five}) and (\ref{nul}))%
\begin{equation}
\mu ^{-1/\varepsilon }=1+\underset{k=0}{\overset{\infty }{\sum }}%
E_{k}\lambda ^{-(k+1)},  \label{sum}
\end{equation}%
where $\lambda \rightarrow \infty $. The hydrodynamic chain (\ref{zm}) is
the same as the hydrodynamic chain (\ref{min}), because these two chains are
connected by invertible transformation (see (\ref{next}) and (\ref{null}))%
\begin{equation*}
\mu ^{-1/\varepsilon }=1-\underset{k=1}{\overset{\infty }{\sum }}%
E_{-k}\lambda ^{k-1},
\end{equation*}%
where $\lambda \rightarrow 0$.
\end{remark}

Thus, our approach is the following: we mark $N$ arbitrary points $\lambda
=\gamma _{\alpha }$ ($N$ distinct punctures) on the Riemann surface $%
F(\lambda $, $\mu )=0$ (see (\ref{first}) and cf. (\ref{Rim})); then we
obtain special set of coordinates%
\begin{equation}
\mu _{\alpha }=\exp \underset{k=1}{\overset{N}{\sum }}\underset{\gamma
_{\alpha }}{\overset{r^{k}}{\int }}\frac{d\psi _{k}(\tilde{\lambda})}{\gamma
_{\alpha }-f_{k}(\tilde{\lambda})},  \label{gena}
\end{equation}%
which in fact is \textit{fundamental basic of linearly independent solutions}
for the corresponding linear system (\ref{third}). It means that any
solution of linear system (\ref{third}) can be presented as a linear
combination of the basic solutions (\ref{gena}) with some coefficients.
Finally, we just mention, that any over-determined linear system like (\ref%
{third}) must have a general solution which depends on $N$ arbitrary
parameters $\gamma _{\alpha }$. In our case, we should take $N$ infinite
serieses of the conservation law densities $\mu _{\alpha ,k}$ ($k=1$, $2$,
...) starting near already fixed punctures $\gamma _{\alpha }$%
\begin{equation*}
\mu ^{(\alpha )}=\mu _{\alpha }+(\lambda -\gamma _{\alpha })\mu _{\alpha
,1}+(\lambda -\gamma _{\alpha })^{2}\mu _{\alpha ,2}+...\text{, \ \ \ \ \ \ }%
\lambda \rightarrow \gamma _{\alpha }\text{, \ \ \ \ }\alpha =1\text{, }2%
\text{, ... , }N.
\end{equation*}%
Thus, the general solution of linear system (\ref{third}) is%
\begin{equation*}
h(\mathbf{r})=\underset{\beta =1}{\overset{N}{\sum }}\underset{\gamma
_{\beta }}{\overset{r^{\beta }}{\int }}\varphi _{\beta }(\lambda )\mu
^{(\beta )}(\lambda )d\lambda ,
\end{equation*}%
where $\varphi _{\beta }(\lambda )$ are arbitrary functions, and the general
solution of the hydrodynamic type system is given in an implicit form (see (%
\ref{red}), (\ref{blue}), (\ref{minna}), (\ref{pin}) and (\ref{hod})):%
\begin{equation}
\begin{array}{c}
x+[f_{i}(r^{i})-\underset{k=1}{\overset{N}{\sum }}\psi _{k}(r^{k})]t_{1}+%
\frac{1}{f_{i}(r^{i})}\exp [\underset{m=1}{\overset{N}{\sum }}\overset{r^{m}}%
{\int }\frac{d\psi _{m}(\lambda )}{f_{m}(\lambda )}]t_{-1} \\
\\
=\frac{1}{\psi _{i}^{\prime }(r^{i})}\partial _{i}\left[ \underset{\beta =1}{%
\overset{N}{\sum }}\underset{\gamma _{\beta }}{\overset{r^{\beta }}{\int }}%
\varphi _{\beta }(\lambda )\tilde{\mu}^{(\beta )}(\lambda )d\lambda \right] ,%
\end{array}
\label{full}
\end{equation}%
where%
\begin{eqnarray*}
\tilde{\mu}_{\alpha } &=&\exp \left[ -\underset{k=1}{\overset{N}{\sum }}%
\underset{\gamma _{\alpha }}{\overset{r^{k}}{\int }}\frac{d\psi _{k}(\tilde{%
\lambda})}{\gamma _{\alpha }-f_{k}(\tilde{\lambda})}\right] \text{, \ \ \ \
\ \ }\tilde{\mu}(\mathbf{r},\lambda )=\exp \left[ -\underset{k=1}{\overset{N}%
{\sum }}\overset{r^{k}}{\int }\frac{d\psi _{k}(\tilde{\lambda})}{\lambda
-f_{k}(\tilde{\lambda})}\right] , \\
&& \\
\tilde{\mu}^{(\alpha )} &=&\tilde{\mu}_{\alpha }+(\lambda -\gamma _{\alpha })%
\tilde{\mu}_{\alpha ,1}+(\lambda -\gamma _{\alpha })^{2}\tilde{\mu}_{\alpha
,2}+...\text{, \ \ \ \ \ \ }\lambda \rightarrow \gamma _{\alpha }\text{, \ \
\ \ }\alpha =1\text{, }2\text{, ... , }N.
\end{eqnarray*}%
The general solution of a linear system (\ref{third}) can be presented in
most possible explicit form in special case, when values $\varepsilon $ are
\textit{integers} for ''$\varepsilon -$systems''. The case $N=2$ (namely
Euler-Darboux-Poisson equation) was completely investigated (see, for
instance, \cite{Rozh}). Its generalization on $N-$component case (\ref{Pois}%
), or moreover on case of arbitrary \textit{integers} $\varepsilon _{m}$
(see (\ref{ten}))%
\begin{equation}
\partial _{ik}h=\frac{1}{r^{i}-r^{k}}[\varepsilon _{k}\partial
_{i}h-\varepsilon _{i}\partial _{k}h]\text{, \ \ \ \ \ \ }i\neq k.
\label{mina}
\end{equation}%
can be made in the same way as in \cite{Rozh}. For simplicity, here we shall
restrict our consideration on case of (\ref{Pois}) (see \cite{Maks}).

The general solution of (\ref{Pois}) (if $\varepsilon =\pm n$, \ \ $n=1$, $2$%
, ...) is%
\begin{equation*}
h_{(n)}=\underset{k=1}{\overset{N}{\sum }}\frac{d^{n}}{d(r^{k})^{n}}\left[
\frac{\varphi _{k}(r^{k})}{\underset{m\neq k}{\sum }(r^{k}-r^{m})^{n}}\right]
\text{, \ \ \ \ }h_{(-n)}=\underset{k=1}{\overset{N}{\sum }}\overset{r^{k}}{%
\int }\varphi _{k}(\lambda )\underset{m=1}{\overset{N}{\prod }}(\lambda
-r^{m})^{n}d\lambda ,
\end{equation*}%
where $\varphi _{k}(r^{k})$ are $N$ arbitrary functions of a single variable
(if we replace $\varphi _{k}(\lambda )\rightarrow \varphi
_{k}^{(nN+1)}(\lambda )$ in second \textit{negative} case, then all
integrals can be expressed via finite number of derivatives only). Thus,
indeed, these solutions are general (see \cite{Tsar}) for \textit{positive}
and \textit{negative integers} $\varepsilon $ (see (\ref{one}) and (\ref%
{Pois})). The general solutions for (\ref{mina}) can be obtained by
recursive application of Laplace transformations (see \cite{Ferl}) to above
formulas (also, see \cite{Aks}).

If $\varepsilon $ is negative and $\varepsilon \neq -n$, \ \ $n=1$, $2$, ...
, then above-mentioned solution (right) easily to generalize%
\begin{equation*}
h_{\varepsilon }=\underset{k=1}{\overset{N}{\sum }}\underset{\gamma _{k}}{%
\overset{r^{k}}{\int }}\varphi _{k}(\lambda )\underset{m=1}{\overset{N}{%
\prod }}(\lambda -r^{m})^{-\varepsilon }d\lambda ,
\end{equation*}%
where $\gamma _{k}$ ($k=1$, $2$, ... , $N$) are arbitrary constants. If $%
\varepsilon $ is positive and $\varepsilon \neq n$, \ \ $n=1$, $2$, ... ,
then above-mentioned solution (left) easily to generalize just in case, when
$\varepsilon N$ is integer, then%
\begin{equation*}
h_{\varepsilon }=\underset{k=1}{\overset{N}{\sum }}\underset{C_{k}}{\oint }%
\frac{\varphi _{k}(\lambda )d\lambda }{\underset{m=1}{\overset{N}{\prod }}%
(\lambda -r^{m})^{\varepsilon }},
\end{equation*}%
where $C_{k}$ ($k=1$, $2$, ... , $N$) are simple small contours surrounding
the points $\lambda =r^{k}$ ($k=1$, $2$, ... , $N$). However, in general
case (when $\varepsilon $ is positive and $\varepsilon N$ is not integer)
these contours $C_{k}$ could not be closed on corresponding Riemann surface,
because a sum of all phase shifts (for every point) will not be proportional
to $2\pi M$, where $M$ is some integer. For avoiding this problem one can
introduce another set of contours-dumb-bell shaped figures $C_{k,\text{ }%
k+1} $ surrounding every two neighbor points $\lambda =r^{k}$ and $\lambda
=r^{k+1}$ ($k=1$, $2$, ... , $N$). So, integration must change sign twice
from clockwise to anticlockwise, then every time phase shift will be $2\pi $
exactly. However, the number of contours must be equal to $N-1$, because in
opposite case (if number is $N$) all contours became linearly dependent.
Thus, in this general case a general solution of (\ref{Pois}) parametrized
by $N$ arbitrary functions of a single variable is%
\begin{equation*}
h_{\varepsilon }=\underset{k=1}{\overset{N-1}{\sum }}\underset{C_{k,\text{ }%
k+1}}{\oint }\frac{\varphi _{k}(\lambda )d\lambda }{\underset{m=1}{\overset{N%
}{\prod }}(\lambda -r^{m})^{\varepsilon }}+\underset{-\infty }{\overset{0}{%
\int }}\frac{\varphi _{N}(\lambda )d\lambda }{\underset{m=1}{\overset{N}{%
\prod }}(\lambda -r^{m})^{\varepsilon }},
\end{equation*}%
where for simplicity we assume (without lost of generacy), that real parts
of Riemann invariants $r^{k}$ (branch points on a complex Riemann surface)
are numerated as follows: $0<\func{Re}r^{1}<\func{Re}r^{2}<...<\func{Re}%
r^{N}$.

\section{(2+1)-integrable hydrodynamic type systems}

Benney momentum chain (\ref{1}) is equivalent to the hierarchy of ($2+1$)
hydrodynamic type systems embedded in dispersionless KP hierarchy as
Khohlov-Zabolotskaya equation%
\begin{equation}
(u_{t_{2}}-uu_{x})_{x}=u_{t_{1}t_{1}},  \label{hoh}
\end{equation}%
which can be obtained from the coupled equations of the Benney momentum
chain (\ref{1}) and one equation of its first nontrivial commuting flow (see
the next section)%
\begin{equation}
\partial _{t_{2}}A_{k}=\partial
_{x}A_{k+2}+A_{0}A_{k,x}+(k+1)A_{k}A_{0,x}+kA_{k-1}A_{1,x}\text{, \ \ \ \ \ }%
k=0\text{, }1\text{, }2\text{, ...}  \label{sec}
\end{equation}%
by eliminating moments $A_{1}$ and $A_{2}$:%
\begin{equation*}
\partial _{t_{1}}A_{0}=\partial _{x}A_{1}\text{, \ \ \ \ \ \ }\partial
_{t_{1}}A_{1}=\partial _{x}[A_{2}+\frac{1}{2}A_{0}^{2}]\text{, \ \ \ \ \ \ }%
\partial _{t_{2}}A_{0}=\partial _{x}[A_{2}+A_{0}^{2}],
\end{equation*}%
where $u=A_{0}$.

Let us start now with the hydrodynamic chains (\ref{four}) and (\ref{second}%
), eliminate field variable $a_{3}$ from couple equations from first
hydrodynamic chain (\ref{four}) and one equation from the second
hydrodynamic chain (\ref{sec})%
\begin{equation*}
\partial _{t_{1}}a_{1}=\partial _{x}[a_{2}-a_{1}^{2}]\text{, \ \ \ \ \ }%
\partial _{t_{1}}a_{2}=\partial _{x}[a_{3}-a_{1}a_{2}]\text{, \ \ \ \ \ \ }%
\partial _{t_{2}}a_{1}=\partial _{x}[a_{3}-2a_{1}a_{2}+a_{1}^{3}].
\end{equation*}%
Then we came to a new integrable ($2+1$) hydrodynamic type system%
\begin{equation}
u_{t_{1}}=w_{x}\text{, \ \ \ \ \ \ }u_{t_{2}}=w_{t_{1}}+uw_{x}-wu_{x},
\label{news}
\end{equation}%
where%
\begin{equation*}
u=a_{1}\text{, \ \ \ \ }w=a_{2}-a_{1}^{2}.
\end{equation*}%
It is easy to check, that all possible hydrodynamic reductions of this
system (\ref{news}) (see, for instance, approach in \cite{Korot})%
\begin{equation*}
r_{t_{1}}^{i}=\mu _{i}(\mathbf{r})r_{x}^{i}\text{, \ \ \ \ \ \ }%
r_{t_{2}}^{i}=\zeta _{i}(\mathbf{r})r_{x}^{i}\text{, \ \ \ \ \ \ \ }i=1\text{%
, }2\text{, ... }N
\end{equation*}%
are completely the same as those found already (see (\ref{blue})), where%
\begin{equation*}
\zeta _{i}=f_{i}^{2}(r^{i})-f_{i}(r^{i})\underset{k=0}{\overset{N}{\sum }}%
\psi _{m}(r^{m})+\frac{1}{2}\left( \underset{k=0}{\overset{N}{\sum }}\psi
_{m}(r^{m})\right) ^{2}-\underset{k=0}{\overset{N}{\sum }}\overset{r^{k}}{%
\int }f_{k}(\lambda )d\psi _{k}(\lambda ).
\end{equation*}%
Moreover, one can obtain a whole hierarchy of such integrable ($2+1$)
hydrodynamic type systems like (\ref{news}) by eliminating some other field
variables $a_{k}$ in combination with another commuting flows of
hydrodynamic chain (\ref{four}). For example, two another equations (see (%
\ref{tri}) and (\ref{minus}))%
\begin{equation}
\partial _{t_{1}}e^{-c_{0}}=\partial _{x}[-c_{1}e^{-c_{0}}]\text{, \ \ \ \ }%
\partial _{t_{-1}}c_{1}=\partial _{x}e^{c_{0}}  \label{bb}
\end{equation}%
yield a new integrable ($2+1$) hydrodynamic type system (its ($1+1$)
hydrodynamic reductions are exactly).

\section{Another hydrodynamic chain}

Now we start with the integrable hydrodynamic type system \cite{Mok}%
\begin{equation}
r_{t}^{i}=\left[ \underset{m=1}{\overset{N}{\sum }}\varepsilon
_{m}r^{m}-\varepsilon _{i}\underset{m=1}{\overset{N}{\sum }}r^{m}\right]
r_{x}^{i}\text{, \ \ \ \ \ \ }i=1\text{, }2\text{, ... , }N,  \label{mok}
\end{equation}%
when $\varepsilon _{k}$ are arbitrary constants. This system can be
rewritten as the hydrodynamic chain%
\begin{equation}
\partial _{t}c_{k}=c_{1}\partial _{x}c_{k}-c_{0}\partial _{x}c_{k+1}\text{,
\ \ \ \ \ \ \ }k=0\text{, }\pm 1\text{, }\pm 2\text{, ...,}  \label{mokh}
\end{equation}%
where moments%
\begin{equation*}
c_{k}=\underset{m=1}{\overset{N}{\sum }}r^{m}(\varepsilon _{m})^{k}.
\end{equation*}

\begin{theorem}
Under the reciprocal transformation%
\begin{equation*}
dz=\frac{1}{c_{0}}dx+\frac{c_{1}}{c_{0}}dt\text{, \ \ \ \ \ }dy=dt
\end{equation*}%
this hydrodynamic chain linearizes%
\begin{equation}
\partial _{y}c_{k}+\partial _{z}c_{k+1}=0\text{, \ \ \ \ \ }k=0\text{, }\pm 1%
\text{, }\pm 2\text{, ...}  \label{last}
\end{equation}
\end{theorem}

It means that any reductions such as (\ref{mok}) of the hydrodynamic chain (%
\ref{mokh}) linearizes under above reciprocal transformation. The solution
of the hydrodynamic chain (\ref{last}) is a set of the separate
Riemann-Monge-Hopf equations%
\begin{equation*}
r_{y}^{i}+f_{i}(r^{i})r_{z}^{i}=0\text{, \ \ \ \ \ }i=1\text{, }2\text{, ...
, }N,
\end{equation*}%
where $f_{i}(r^{i})$ are arbitrary functions. Thus, every integrable
reduction of hydrodynamic chain (\ref{mokh}) has the simple form%
\begin{equation*}
r_{t}^{i}=[c_{0}f_{i}(r^{i})-c_{1}]r_{x}^{i},
\end{equation*}%
where%
\begin{equation*}
c_{k}=\underset{m=1}{\overset{N}{\sum }}\overset{r^{m}}{\int }[f_{m}(\lambda
)]^{k}d\psi _{m}(\lambda )\text{, \ \ \ \ }k=0\text{, }\pm 1\text{, }\pm 2%
\text{, ...}
\end{equation*}%
and $\psi _{m}(r^{m})$ are arbitrary functions (by scaling $\psi
_{m}(r^{m})\rightarrow R_{m}$ integrable hydrodynamic reductions are
parameterized by $N$ arbitrary functions of a single variable only).

\section{Open Problems}

The Benney momentum chain (the Zakharov reduction) is a dispersionless limit
of the vector nonlinear Shrodinger equation (see (\ref{VNLS}), (\ref{Rim})
and \cite{Zakh}). The inverse problem is: how to reconstruct a \textit{%
dispersive} integrable analogue of given hydrodynamic type system. A \textit{%
dispersive} analogue is known (Coupled KdV is a \textit{dispersive} analogue
of system (\ref{one}); Couple Harry Dym is a \textit{dispersive} analogue of
system (\ref{com}), see, i.g. \cite{Fordy} and \cite{Fer}) just in case of ''%
$\varepsilon -$systems'' with $\varepsilon =-1/2$. The KP hierarchy is a
\textit{dispersive} analogue for the \textit{whole} Benney momentum chain (%
\ref{1}) (as KP equation is a \textit{dispersive} analogue of the
Khohlov-Zabolotzkaya system (\ref{hoh})), but similar \textit{dispersive} ($%
2+1$) analogues for the whole hydrodynamic chain (\ref{tri}) or for ($2+1$)
hydrodynamic type systems (\ref{news}) or (\ref{bb}) still are unknown.

Local Hamiltonian structures for the hydrodynamic type system (\ref{one})
were completely investigated in \cite{Maks3} and \cite{Fer}. It was proved,
that if $N=2$, then the hydrodynamic type system (\ref{one}) for any $%
\varepsilon $ has three local Hamiltonian structures (also, see \cite{Nutku}
and \cite{Ferr}); if $N=3$ and $\varepsilon =-1/2$, then it has four local
Hamiltonian structures; if $N=3$ and $\varepsilon =1$, then it has
two-parametric family of local Hamiltonian structures (also see \cite{Maks4}
and \cite{Fer+Maks}); if $N>3$, then $\varepsilon =-1/2$ and it has ($N+1$%
)-- local Hamiltonian structures.

Hamiltonian structures of integrable hydrodynamic type systems are
determined by a metric $g_{ii}$ (see, \cite{Dubn}). The metric%
\begin{equation*}
g^{ii}=\zeta _{i}(r^{i})\exp \left[ -2\underset{k\neq i}{\sum }\overset{r^{k}%
}{\int }\frac{d\psi _{k}(\lambda )}{f_{i}(r^{i})-f_{k}(\lambda )}\right]
\end{equation*}%
with an arbitrary functions $\zeta _{i}(r^{i})$ determine a nonlocal
Hamiltonian formalism (see \cite{FerapFunk} and \cite{Tsar}) of hydrodynamic
type systems (\ref{red}), (\ref{blue}), (\ref{minna}), (\ref{pin}) and their
commuting flows. Unfortunately, local and nonlocal Hamiltonian formalism has
been done just for the hydrodynamic type systems (\ref{one}) when $%
\varepsilon =\pm 1$ and $\varepsilon =-1/2$ (see for instance \cite{Fer+Maks}
and \cite{FerapFunk}). However, the problem of a description of local and
nonlocal Hamiltonian structures in general case (\ref{blue}) still is open.
Nevertheless, this problem can be solved by the Dirac restriction of a
Hamiltonian structure (see for the beginning \cite{FerapFunk}) known for the
whole hydrodynamic chain, as it was already done in \cite{Blaszak} for
another hydrodynamic chains.

Starting point of such investigation is a Lax-like representation. For
instance, the Lax-like representation (see (\ref{2})) for the dispersionless
KP hierarchy (i.e. the Benney momentum chain (\ref{1})) is well known (see %
\cite{Kodama})%
\begin{equation}
\partial _{t_{n}}\lambda =\{Q_{n}\text{, }\lambda \}=\frac{\partial Q_{n}}{%
\partial \mu }\frac{\partial \lambda }{\partial x}-\frac{\partial Q_{n}}{%
\partial x}\frac{\partial \lambda }{\partial \mu }\text{, \ \ \ \ }n=0\text{%
, }1\text{, }2\text{, ... ,}  \label{Lax}
\end{equation}%
where $Q_{n}$ is the part, polynomial in $\mu $, of $\lambda ^{n}$. Also,
the first local Hamiltonian structure for whole Benney momentum chain (\ref%
{1})%
\begin{equation}
\partial _{t_{n}}A_{k}=\underset{m\geqslant 0}{\sum }[kA_{k+m-1}\partial _{x}%
\frac{\delta H_{n+1}}{\delta A_{m}}+(mA_{k+m-1}\frac{\delta H_{n+1}}{\delta
A_{m}})_{x}],  \label{Hama}
\end{equation}%
where the Hamiltonian is $H_{2}=\frac{1}{2}\int [A_{2}+A_{0}^{2}]dx$, was
constructed in \cite{Manin} (the relationship between formulas (\ref{Lax})
and (\ref{Hama}) was found in \cite{Kodama} too; first notrivial commuting
flow (see (\ref{sec})) is determined by the next Hamiltonian $H_{3}=\frac{1}{%
3}\int [A_{3}+3A_{0}A_{1}]dx$; functional $H_{0}=\int A_{0}dx$ is a Casimir
of this Hamiltonian structure, the functional $H_{1}=\int A_{1}dx$ is a
momentum of this Hamiltonian structure).

Similar Lax-like representation for the hydrodynamic chain (\ref{tri}) was
established in \cite{Shabat} (cf. (\ref{Lax}))%
\begin{equation}
\partial _{t_{n}}L=\left\langle Q_{n}\text{, }L\right\rangle =Q_{n}\frac{%
\partial L}{\partial x}-\frac{\partial Q_{n}}{\partial x}L\text{, \ \ \ \ \ }%
n=0\text{, }1\text{, }2\text{, ... ,}  \label{Laxa}
\end{equation}%
where%
\begin{equation*}
Q_{n}=(\lambda ^{n}L)_{+}\text{, \ \ \ \ \ \ \ \ }L=1+G_{0}/\lambda
+G_{1}/\lambda ^{2}+G_{2}/\lambda ^{3}+...
\end{equation*}%
The corresponding first hydrodynamic chain%
\begin{equation*}
\partial _{t_{1}}G_{k}=\partial _{x}G_{k+1}+G_{0}G_{k,x}-G_{k}G_{0,x}\text{,
\ \ \ \ \ \ }k=0\text{, }1\text{, }2\text{, ... }
\end{equation*}%
is exactly the hydrodynamic chain (\ref{z}), where $\varepsilon =1$, and
linear term $-\varepsilon \left( \underset{\beta =1}{\overset{N}{\sum }}%
\gamma _{\beta }\right) E_{k,x}$ is removed by shift of independent variable
($x\rightarrow x-\varepsilon \left( \underset{\beta =1}{\overset{N}{\sum }}%
\gamma _{\beta }\right) t$). Thus, a generating function of these moments is
(see (\ref{z}), (\ref{sum}) and (\ref{Laxa}))%
\begin{equation*}
\mu ^{-1}\equiv L=1+\underset{k=0}{\overset{\infty }{\sum }}G_{k}\lambda
^{-(k+1)}.
\end{equation*}%
The alternative Lax-like representations are%
\begin{equation*}
\partial _{t_{k}}\rho =[\underset{m=0}{\overset{k}{\sum }}\tilde{a}%
_{m}\lambda ^{k-m}\partial _{x}\text{, }\rho ]\text{, \ \ \ \ \ }\partial
_{t_{-k-1}}\rho =[\underset{m=0}{\overset{k}{\sum }}\tilde{b}_{m}\lambda
^{m-k-1}\partial _{x}\text{, }\rho ]\text{, \ \ \ \ \ }k=0\text{, }1\text{, }%
2\text{, ... ,}
\end{equation*}%
where $\mu =\rho _{x}$ (see (\ref{ra}), (\ref{pa})).

We suppose that the hydrodynamic chain (\ref{tri}) and its commuting flows
have local Hamiltonian structure (sf. (\ref{Hama}))%
\begin{equation*}
\partial _{t_{n}}c_{k}=\underset{k=1}{\overset{\infty }{\sum }}[\beta _{k,m}(%
\mathbf{c})\partial _{x}+\partial _{x}\beta _{m,k}(\mathbf{c})]\frac{\delta
H_{n+1}}{\delta c_{m}},
\end{equation*}%
where $\beta _{k,m}$ are some functions.

The Hamiltonian structures of integrable hydrodynamic type systems can be
successfully investigated by application of methods from the differential
(see, for example, \cite{Tsar}, \cite{Maks}--\cite{Maks4}, \cite{Ferr}, \cite%
{Fer}, \cite{FerapFunk} and \cite{Fer+Maks}) algebraic geometry (see, for
instance, \cite{Dubr} and \cite{Krich}). An alternative way is following:
assume that our given integrable hydrodynamic type system ($N$ components)
is a some reduction of some ''bigger'' integrable hydrodynamic type system ($%
N+M$ components); assume that Hamiltonian structure of such ''bigger''
integrable hydrodynamic type system is already known. Then the direct
application of the Dirac restriction to this Hamiltonian structure (the
first step in such procedure (see \cite{FerapFunk}) is the choice of some
Riemann invariants $r^{k}=\func{const}$, $k=1$, $2$, ... , $M$) yields the
transformed Hamiltonian structure of a ''restricted'' hydrodynamic type
system. The Dirac restriction of Hamiltonian structures (in algebraic
language) was developed in application to hydrodynamic chains and their
reductions (see \cite{Blaszak}). The first step in such procedure is the
recalculation of the Lax-like representation (such (\ref{Lax}) or (\ref{Laxa}%
)) to the Hamiltonian structure of a whole hydrodynamic chain (see for
instance (\ref{Hama})).

\section{Conclusion}

In this article we present a recipe: how to construct a hydrodynamic chain
starting from any given hydrodynamic type system with \textit{polynomial}
(or \textit{rational}) velocities with respect to their field variables (for
simplicity we have mentioned just two cases: namely Benney momentum chain,
whose moments are connected directly with some \textit{conservation law
densities} ($u_{k}$, $\eta _{k}$) and the hydrodynamic chain (\ref{tri}),
whose moments are connected directly with \textit{Riemann invariants}). In
fact, it means that \textbf{any integrable hydrodynamic type system (written
in Riemann invariants) with such \textit{polynomial} velocities is embedded
in hydrodynamic chain} (\ref{tri}) \textbf{or its higher (or lower)
commuting flows.} Any integrable hydrodynamic type system written in Riemann
invariants with \textit{rational} velocities%
\begin{equation}
r_{t}^{i}=g_{0}\frac{(r^{i})^{M}+g_{1}(r^{i})^{M-1}+...+g_{M}}{%
(r^{i})^{K}+e_{1}(r^{i})^{K-1}+...+e_{K}}r_{x}^{i}\text{, \ \ \ \ \ \ \ }i=1%
\text{, }2\text{, ..., }N  \label{*}
\end{equation}%
has a generating function of conservation laws%
\begin{equation}
\mu _{t}=\partial _{x}[g_{0}\frac{\lambda ^{M}+g_{1}\lambda ^{M-1}+...+g_{M}%
}{\lambda ^{K}+e_{1}\lambda ^{K-1}+...+e_{K}}\mu ].  \label{**}
\end{equation}%
For simplicity we assume that coefficients $e_{k}$ and $g_{k}$ of such
\textit{rational} velocities are some \textit{symmetric }(not necessary to
be \textit{polynomials}!) functions of Riemann invariants; $K$, $M$ and $N$
are arbitrary natural numbers. As example, we can take any integrable
systems embedded into $2x2$ spectral transform like Korteweg de Vries
equation, Bonnet equation (Sin-Gordon equation) and nonlinear Shrodinger
equation. All their Whitham deformations (i.e. hydrodynamic type systems,
see, for instance, \cite{Dubr}, \cite{Krich}) have such representation (\ref%
{*}) as consequence (\ref{**}) written in abelian differentials on
hyperelliptic surfaces%
\begin{equation*}
\partial _{t}dp=\partial _{x}dq,
\end{equation*}%
where%
\begin{equation*}
dp=\frac{\lambda ^{K}+e_{1}\lambda ^{K-1}+...+e_{K}}{\sqrt{\overset{N}{%
\underset{m=1}{\prod }}(r_{m}-\lambda )}}d\lambda \text{, \ \ \ \ \ \ \ }%
dq=g_{0}\frac{\lambda ^{M}+g_{1}\lambda ^{M-1}+...+g_{M}}{\sqrt{\overset{N}{%
\underset{m=1}{\prod }}(r_{m}-\lambda )}}d\lambda ,
\end{equation*}%
and velocities of (\ref{*}) are%
\begin{equation*}
g_{0}\frac{(r^{i})^{M}+g_{1}(r^{i})^{M-1}+...+g_{M}}{%
(r^{i})^{K}+e_{1}(r^{i})^{K-1}+...+e_{K}}=\frac{dq}{dp}\mid _{\lambda
=r^{i}}.
\end{equation*}%
Substituting (sf. (\ref{nul}))%
\begin{equation*}
\mu =1+\overset{\infty }{\underset{k=1}{\sum }}a_{k}\lambda ^{-k}
\end{equation*}%
into generating function (\ref{**}) one can obtain similar formulas and
results as it was done in this paper. Next step is \textit{replication} of
integrable hydrodynamic type systems as different hydrodynamic reductions of
hydrodynamic chains. The \textit{main advantage} of such \textit{replicated}
systems is \textit{preservation} of some properties of original hydrodynamic
systems like \textit{generating functions of conservation laws and commuting
flows} (see (\ref{2}), (\ref{4}) and (\ref{6}) for Benney momentum chain (%
\ref{1}); see (\ref{six}), (\ref{five}) and (\ref{zero}) for the
hydrodynamic chain (\ref{tri})). Thus, a problem of integrability is much
simpler -- all that necessary to do: to construct a general solution
(starting from already obtained generating function, see (\ref{7}), (\ref%
{hod}) and (\ref{full})), parameterized by $N$ functions of a single
variable (see \cite{Tsar}) and to solve Cauchy problem, that in fact is done
to this moment just for four cases (the Zakharov reduction of the Benney
momentum chain, see \cite{Geog}; linearly degenerate system, see \cite{Tsarr}%
, this particular class is ''$\varepsilon -$systems'' with $\varepsilon =1$;
hydrodynamic type systems of the Tample class, this particular class is ''$%
\varepsilon -$systems'' with $\varepsilon =-1$; Whitham hydrodynamic type
systems related with hyperelliptic surfaces such averaged $N-$phase
solutions of Korteweg de Vries equation (KdV)\ or nonlinear Shrodinger
equation (NLS), it was done in articles of \textit{G. El, T. Grava, B.A.
Dubrovin, F.R. Tian, J. Gibbons} and many others).

Moreover, any two commuting flows of (\ref{five}), for example (\ref{ra})%
\begin{equation*}
dz=\mu \lbrack dx+\underset{m=0}{\overset{k}{\sum }}\tilde{a}_{m}\lambda
^{k-m}dt^{k}+\underset{m=0}{\overset{n}{\sum }}\tilde{a}_{m}\lambda
^{n-m}dt^{n}],\ \ \ \ \ k,n=1,2,...
\end{equation*}%
yield hydrodynamic type systems with \textit{rational} velocities%
\begin{equation*}
r_{t_{k}}^{i}=\frac{\underset{m=0}{\overset{k}{\sum }}\tilde{a}%
_{m}(r^{i})^{k-m}}{\underset{m=0}{\overset{n}{\sum }}\tilde{a}%
_{m}(r^{i})^{n-m}}r_{t_{n}}^{i}\text{, \ \ \ \ \ }k\neq n\text{, \ \ }{\ \ \
}i=1\text{, }2\text{, ..., }N.
\end{equation*}%
A more complicated \textit{rational} dependence can be obtained by
application of a generalized reciprocal transformation (see, for instance, %
\cite{Ferap} and \cite{Fer+Maks}), starting from (\ref{five}) and its
commuting flows.

Thus, this is \textit{powerful tool for classification} of integrable
hydrodynamic type systems and their \textit{integrability}. Moreover, if any
given hydrodynamic type system has a generating function of conservation
laws (see for instance, (\ref{4}) or (\ref{five})), it means that the
corresponding hydrodynamic chain has the same generating function. For
example, if a some hydrodynamic type system has the same generating function
as the Benney momentum chain (\ref{4}), it means that this hydrodynamic type
system is a some reduction of the Benney momentum chain. Thus, this is a
wonderful symptom in recognition of an immersion of any unknown hydrodynamic
type systems into already known hydrodynamic chains. Thus, \textit{if one
can construct a generating function of conservation laws for some
hydrodynamic type system, it means that, in fact, hydrodynamic chain is
already constructed} (and may be recognized, because, obviously, amount of
hydrodynamic chains is much smaller than amount of integrable hydrodynamic
type systems).

However, the problem of a description of all possible reductions is very
complicated. For instance, this problem for the Benney momentum chain is
still open (see \cite{Gibbons}). However, this problem for the Boyer-Finley
momentum chain (continuum limit of the Darboux-Laplace chain, which also is
known as two-dimensional Toda lattice, see \cite{Sheftel}, \cite{Korot} and %
\cite{Krich}) in fact is \textit{not exist}, because both mentioned
hydrodynamic chains are related by special exchange of independent
variables, see \cite{Kodama1}). Thus, we are lucky to solve this problem for
the hydrodynamic chain (\ref{tri}).

\begin{acknowledgement}
I am grateful to my friends and colleagues Eugeni Ferapontov for his
suggestions in improvement of my calculations and Yuji Kodama for useful
discussions.
\end{acknowledgement}

\end{document}